\documentclass[preprint]{aastex}
\usepackage{psfig}
\def\HI {H\kern0.1em{\sc i}} 
\def\radm {rad m$^{-2}$} 

\def\dg{$^{\circ}$}

\begin{document}
\title{~~\\ ~~\\ Time Variable Faraday Rotation Measures of 3C\,273 \& 3C\,279}
\shorttitle{Variable Rotation Measures in 3C\,273 and 3C\,279}
\shortauthors{Zavala \& Taylor}
\author{R. T. Zavala\altaffilmark{1,2} \& G. B. Taylor\altaffilmark{1}}

\email{rzavala@nmsu.edu, gtaylor@nrao.edu}
\altaffiltext{1}{National Radio Astronomy Observatory, P. O. Box 0, Socorro, NM
  87801, USA}
\altaffiltext{2}{Department of Astronomy, New Mexico State University, Dept. 4500 P. O. Box 30001 
Las Cruces, NM 88003, USA}



\begin{abstract}

Multifrequency polarimetry with the VLBA confirms the previously reported time-varying Faraday
rotation measure (RM) in the quasar 3C\,279. Variability in the RM and electric vector position angle 
(EVPA) of the jet component (C4) is seen making it an unreliable absolute EVPA calibrator. 3C\,273 
is also shown to vary its RM structure on 1.5
year time-scales.     Variation in the RM properties of quasars may result from a Faraday screen which
changes on time-scales of a few years, or from the motion of jet components which sample spatial
variations in the screen. A new component emerging from the core of 3C\,279 appears to be starting to 
sample such a spatial variation. Future monitoring of this component and its RM properties is suggested
as a diagnostic of the narrow line region in 3C\,279. 
We also present a new method of EVPA calibration using the VLA Monitoring Program.

\end{abstract}


\section{Introduction}

The parsec-scale cores and jets of quasars have typical linear polarizations
of 1-10\% \citep{caw93} with the jets being more strongly
polarized.   If a change in the electric vector position
angle (EVPA) of the polarized emission is seen with frequency,
and can be attributed to the Faraday effect, then we can learn
about the densities and magnetic fields of thermal gas along the line
of sight.  

High angular resolution polarimetry has shown that some quasars have Faraday Rotation Measures (RM) in
excess of $\sim 1000 $ \radm . In the case of the quasar 3C\,279
 these high RMs can change over 1.5 years by over 1000 {\radm} (\citet{tay00} and 
references therein). These high RMs are present within a projected distance of $\sim 20$ pc from the
core of the quasars. Beyond this distance the RMs decrease to $\leq 100 $ \radm, a value consistent
with that expected from the passage of the emission through the host galaxy 
and through the ISM of our own Galaxy.

In \S 2 we describe VLBA observations of the quasars 3C\,273 and 3C\,279, with the results of the 
rotation measure analysis presented in \S 3. In \S 4 we discuss our results in light of the unified scheme of AGNs.

We assume H$_0 = 50$ km s$^{-1}$ Mpc$^{-1}$ and q$_0$=0.5 throughout.
  
\section{Observations and Data Reduction}
The observations, performed on 2000 January 27 (2000.07), were carried out at 
ten frequencies between 8.1 and 43.2 GHz 
 using the 10 element VLBA\footnote{The National Radio Astronomy Observatory is operated by 
Associated Universities, Inc., under cooperative agreement with the National 
Science Foundation.}. Sources 3C\,273, 3C\,279, and calibrator J1310+3220 were observed for 
approximately 2.5 hours each. The calibrator OJ287 was observed for almost one hour. 
Right- and left-circular polarizations 
were recorded using 1 bit sampling across a bandwidth of 8 MHz. The 
VLBA  correlator produced 16 frequency channels across each 
IF during every 2 s integration.

Amplitude calibration and fringe-fitting were performed as in \citet{tay00}. Weather, in 
particular snow on the antenna, caused us to flag the data 
from Hancock at 43 GHz. We also encountered radio-frequency interference at 12 GHz on our 
3C\,279 scans, which we attribute
to geostationary television broadcasts. This required extensive editing of the 12 GHz data on 3C\,279. 

Feed polarizations of the antennas were determined using 3C\,279 and the AIPS task LPCAL. We assumed 
that the VLBA antennas had good-quality feeds with relatively pure polarizations, which allowed us to
use a linearized model to fit the feed polarizations. Absolute electric vector position angle (EVPA)
calibration at 8--22 GHz was determined by using the EVPA of OJ287 obtained on 2000 January 29 listed in the 
VLA Monitoring Program\footnote{http://www.aoc.nrao.edu/$\sim$smyers/calibration/} (Taylor \& Myers 2000). Polarization monitoring observations of J1310+3220 on 1999 December 19 
and 2000 February 18 were interpolated to the observation date and employed as a check of the OJ287 calibration (Table 1).  At 43 GHz the low polarized flux density from these calibrators 
prevented an absolute EVPA calibration. To set the EVPA at 43 GHz we performed a linear 
least-squares fit to the 
3C\,279 component C4 EVPA data, and used this fit to set the EVPA for 3C\,279 at 43 GHz. The correction 
applied for 3C\,279 C4 was then used for the other sources (Fig 1). 

\citet{tay98} suggested that C4 may be used as an absolute EVPA calibrator due to its low RM and strong polarized and total flux densities. From 1994 \citep{lep95} through 1998  \citep{lis98,war98,den00,tay00} component C4 maintained an EVPA oriented nearly parallel to the jet axis.  If we define the intrinsic polarization angle (IPA) of a source as the 
intercept in the 
RM plot \citep{sim81} the IPA of 3C\,279 C4 determined by our epoch
2000.07 observations is ${-87.1 \pm 2.2^{\rm o}}$.  The position angle 
of component C4 is significantly different from the 
value of ${\sim -120^{\rm o} \pm 3^{\rm o}}$ 
found by \citet{tay98}. Therefore we conclude that C4 
cannot be used as an absolute EVPA calibrator, without complimentary monitoring observations.

\section{Results}

We present multifrequency radio observations, polarization 
properties, and rotation measure distributions for 3C\,273 and
3C\,279. In the rotation measure images presented in figures 2 through 5
the higher frequencies have been tapered to provide a matched restoring beam 
of the dimensions indicated in the captions. 

\subsection{3C\,273}
Figure 2 presents a time series RM mosaic for this quasar (epoch 1997.07 from 
Taylor 1998). Beyond 10 pc
from the nuclear region the jet  exhibits a relatively
smooth and uniform RM distribution of $\sim$ +500 {\radm}. The RM of this region remains constant over a 3 year 
time span. This jet continues to show good agreement with the ${\lambda^2}$ law of Faraday 
rotation. In contrast the nuclear region changes from an apparently smooth 
$-$2000 {\radm} in 1997 to a more complex structure which varies by ${\mid4000\mid}$ {\radm} 
across 10 pc. To investigate the structure of this complex central region we created
a higher resolution RM map in Figure 3. This map consists of data from 15-43 
GHz with the high frequency data tapered to the resolution of the 15 GHz 
data. Inserted in this figure are the data used to obtain the indicated rotation measures. 
All points used to calculate the RM are at least 1\% polarized.  The central region consists of 
sub-components of RM in the range ${\mid1000-2000\mid}$ {\radm}. Good agreement to a 
${\lambda^2}$ law is present except in the northeastern-most part of the source, closest to the center of activity.

\subsection{3C\,279}
Figure 4 presents a multi-epoch RM map of 3C\,279 (first two epochs from
Taylor 1998 and Taylor 2000 respectively). The core may be ejecting a new jet component, 
as shown by the increasing extension of the emission to the southwest over the three 
epochs. \citet{tay00} previously showed that 3C\,279 shows variation in its RM properties. 
The figure illustrates these changes as the central region of 3C\,279 changes from $-1300$ to 
$>-500$ {\radm} between the 1997.07
 and 1998.5 observations. 3C\,279 maintains this lower RM during our 2000.07 
observation with the exception of a ``bridge'' of high RM which appears to 
connect the central region with component C4 \citep{lep95}.  

The high resolution RM structure of this quasar is presented in Figure 5. The 
rotation measures  were determined using the 12-43 GHz data tapered to the 12 
GHz resolution. The bridge appearing in the low resolution data is resolved 
and two edges of ${\sim}$ 500 {\radm} bordering the central region of the quasar 
and C4 appear. Points used to fit the RM in the center of the image and in component 
C4 are at least 1\% polarized. Points used in the bridge range from 0.3-0.5\% 
polarized intensity and should be interpreted with caution. The asymmetry of the central region is 
consistent with the emergence 
of a new component from the core. This high resolution image also shows the slope
of the RM changing from negative at the central peak of emission to a positive slope
at the proposed newly emerging component.  

\section{Discussion}
Taylor (1998, 2000) showed variation in the RM structure of AGNs on small 
spatial scales. For 3C\,279 \citet{tay00} presented time variation of the RM 
structure of 
3C\,279 over 1.5 years. Our 2000.07 observations 
show temporal variations in 3C\,273 and demonstrate that this phenomenon is 
not unique to 
3C\,279.
We find good agreement to a ${\lambda^2}$ law except near the core of 3C\,273 
and the resolved bridge of 3C\,279. The high RMs observed in the central regions 
could result from the passage of the radiation through the narrow-line region 
(NLR) of the quasar.
If the measured RM results from a polarized component which dominates within 
the telescope beam then 
the RM of the central region should vary as new components are created and 
ejected. As these
new components emerge from the core their radiation  passes through the NLR 
and a high RM 
results. The component progresses along the jet and the RM varies as the line 
of sight 
traverses an 
inhomogeneous NLR Faraday screen. Eventually RMs below 100 {\radm} should be observed, which
are comparable to RMs produced by traversing the ISM of the galaxy, as the NLR is exited. 
With this picture the jet of 3C\,273 may still probe the NLR as the RM in the jet of 
this quasar in the 8-15 GHz 
mosaic is approximately 5 times that observed in radio galaxy hosts \citep{fer99}. We intend
to monitor the cores of quasars for time variations in the 
RM created as new components are ejected and sample these regions.

A crude estimate of the magnetic field strengths in these quasars can be made by using
the variance in the observed RMs \citep{fel96}. This attributes the observed 
RMs to a randomly ordered magnetic field embedded in cells of uniform size 
and field strength. From our high resolution RM images we estimate 
a cell size ($\it{l}$) of 1 pc. We estimate $\sigma$ in these maps as 1160
$\rm rad m^{-2}$ for 3C\,273 and 255 $\rm rad m^{-2}$ for 3C\,279. 
With an electron density $n_{e} = 10^4$ $\rm{cm}^{-3}$, a volume filling factor 
$\epsilon = 0.01$ and a path length $X$ through the NLR of 100 pc the $B$ field 
can be estimated using 
\begin{equation}
B = (3/l)^{1/2}(\sigma/812\epsilon{n_e}X).
\end{equation}
This results in a $B$ field of 0.3 mG in 3C\,273 and 0.05 mG for 3C\,279. 
This is less than that 
estimated by assuming the magnetic and thermal pressures to be in equilibrium
for the same physical conditions \citep{udo97}. This estimate is also less than 
that estimated for the accretion disks of other AGNs obtained by equating 
magnetic to thermal pressures \citep{jon00}. Direct $B$ field measurements for 
other galaxies may be useful in setting a scale of reasonable $B$ field values 
to use in interpreting these quasar RM results. Zeeman splitting measurements 
in the Seyfert galaxy NGC 4258 provide an upper limit on the $B$ field within 
0.2 pc of the central engine of this AGN of 300 mG \citep{her98}. In the more 
benign region of the center of our own galaxy Zeeman measurements yield $B$ 
fields varying from 0.5 mG in Sgr B2 ($\sim$ 120 pc from the galactic center) 
\citep{cru96} to 4 mG within 2.5 pc from Sgr $\rm A^*$ \citep{pla95}. 
It seems reasonable that the field strengths within the NLR of a powerful 
quasar should be at least as strong as that in the center of our own galaxy, 
so the variance method of Felten may provide a lower limit on the $B$ field 
present in the NLR of these quasars.
   

\section{Conclusions}
Both 3C\,273 and  3C\,279 exhibit RM variations on spatial scales of a few parsecs and  
1-3 year temporal scales.
The low resolution RM structure of 3C\,273 is resolved into sub-components whose general 
structure is visible at low resolution. Good agreement to a ${\lambda^2}$ law is found outside 
of the 
core of this quasar.  We estimate a $B$ field of 0.3 mG for 3C\,273 and 0.05 mG for 3C\,279 
if the RMs result from 
a tangled field in 1 pc cells.  Component C4 of 3C\,279 is shown to have a varying 
absolute EVPA and 
thus is not 
suitable as a stand-alone EVPA calibrator. We find that the VLA Monitoring Program database provides
a method of absolute EVPA calibration for VLBA observations \citep{tmy00}. 
The RM slope change from negative to positive in the new component of 3C\,279 
emerging from the core may also indicate that a new positive RM feature in the NLR is
being sampled, and its progress will be examined in future VLBA observations.  

\acknowledgments
This research has made use of the NASA/IPAC Extragalactic Database (NED)
which is operated by the Jet Propulsion Laboratory, Caltech, under
contract with NASA. This research has made use of NASA's Astrophysics
Data System Abstract Service. R.T.Z. gratefully acknowledges support from 
the NRAO Summer Graduate Student Research Program and NSF Grant HRD-9628730.

\clearpage

\begin{center}
TABLE 1 \\
\smallskip
P{\sc olarization} A{\sc ngle} C{\sc alibration}
\smallskip
 
\begin{tabular}{l r r r r r r r r}
\hline
\hline
Source & $\nu$ & $S_{\rm VLA}$ & $P_{\rm VLA}$ & $\chi_{\rm VLA}$ & $S_{\rm VLBA}$ & $P_{\rm VLBA}$
& $\chi_{\rm VLBA}$ \\
(1) & (2) & (3) & (4) & (5) & (6) & (7) & (8) \\
\hline
\noalign{\vskip2pt}
OJ287    & 8.5 & 2.6 & 148 & $-$52.5 & 2.5 & 140 & 26 \\
         & 22  & 2.5 & 100 & -53 & 1.9 & 81 & $-$61 \\
         & 43   & 2.1 & 89 & $-$37 & 1.1 & 8 & $-$88 \\
J1310+3220 & 8.5 & 1.7 & 56 & $-$20 & 1.8 & 52 & 58 \\
         & 22  & 1.2 & 59 & $-$27.7 & 1.0 & 56 & $-$26 \\
         & 43   & 1.0 & 52 & $-$29 & 1.1 & 36 & $-$14 \\
\hline
\end{tabular}
\end{center}
\smallskip
\begin{center}
{\sc Notes to Table 1}
\end{center}
Col.(1).---Source name.
Col.(2).---Observing band in GHz.
Col.(3).---Integrated VLA flux density in Jy.
Col.(4).---Integrated VLA polarized flux density in mJy.
Col.(5).---VLA polarization angle (E-vector) in degrees.
Col.(6).---Integrated VLBA flux density in Jy.
Col.(7).---Integrated VLBA polarized flux density in mJy.
Col.(8).---VLBA polarization angle (E-vector) in degrees.
\bigskip

\clearpage

\begin{figure}
\vspace{19.2cm}
\includegraphics{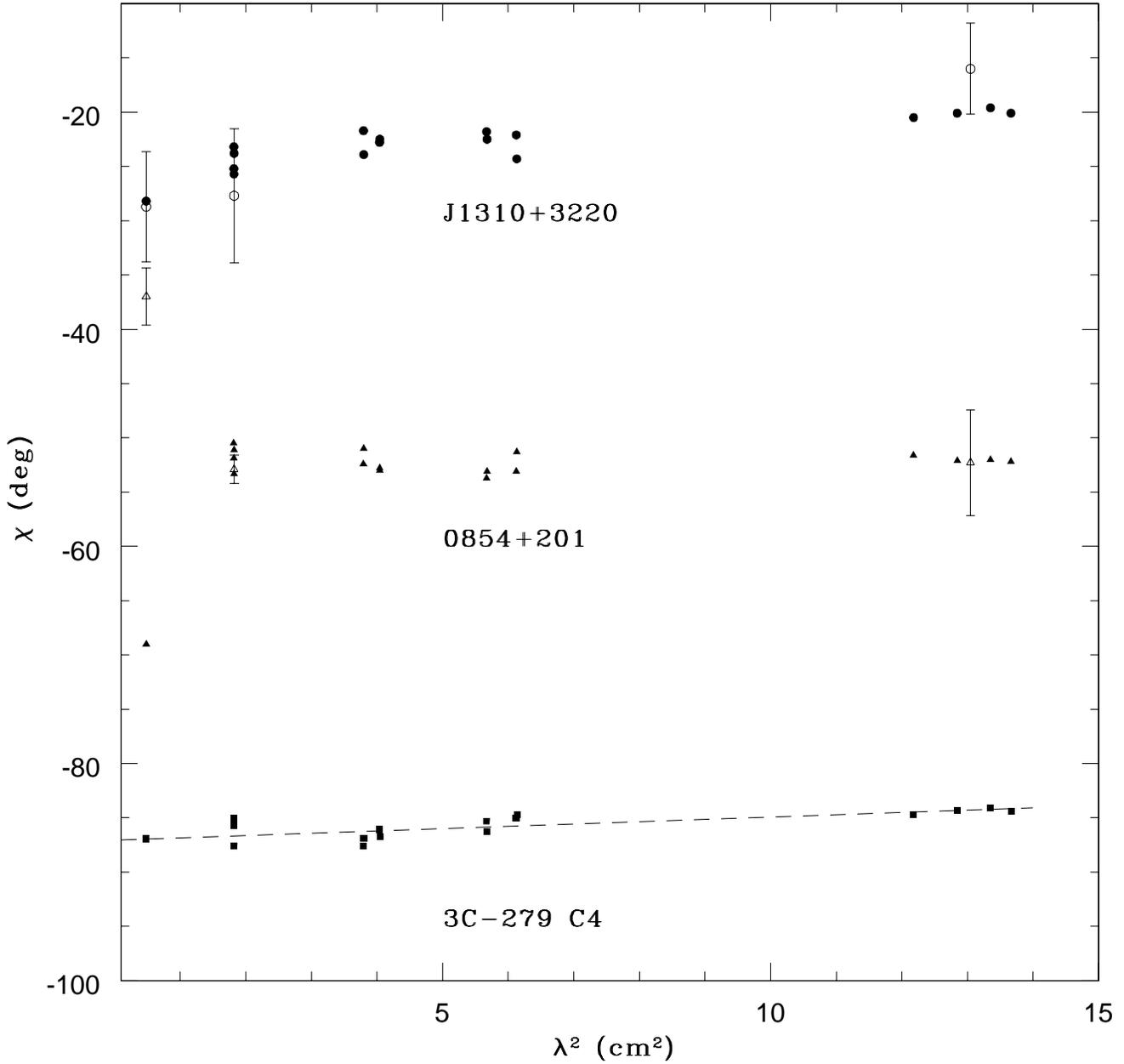}
\figcaption{Plot of EVPA vs. $\lambda^2$ for calibrators J1310+3220 (circles) and
 OJ287 (triangles) and
component C4 of 3C\,279 (squares).  Filled symbols are EVPA after applying
correction based on VLA polarization monitoring data for 8--22GHz. Open
symbols represent VLA polarization monitoring data for the calibrators.
The least squares fit shown (dashed line) for C4 was used to set the EVPA at 43 GHz
for all sources. Uncertainty in position angle is estimated as ${\pm 5^{\rm o}}$ for all frequencies 
after EVPA calibration. Slope of the fit for 3C\,279-C4 equals a RM of 37 \radm.}
\end{figure}
\clearpage

\begin{figure}
\vspace{19.2cm}
\includegraphics{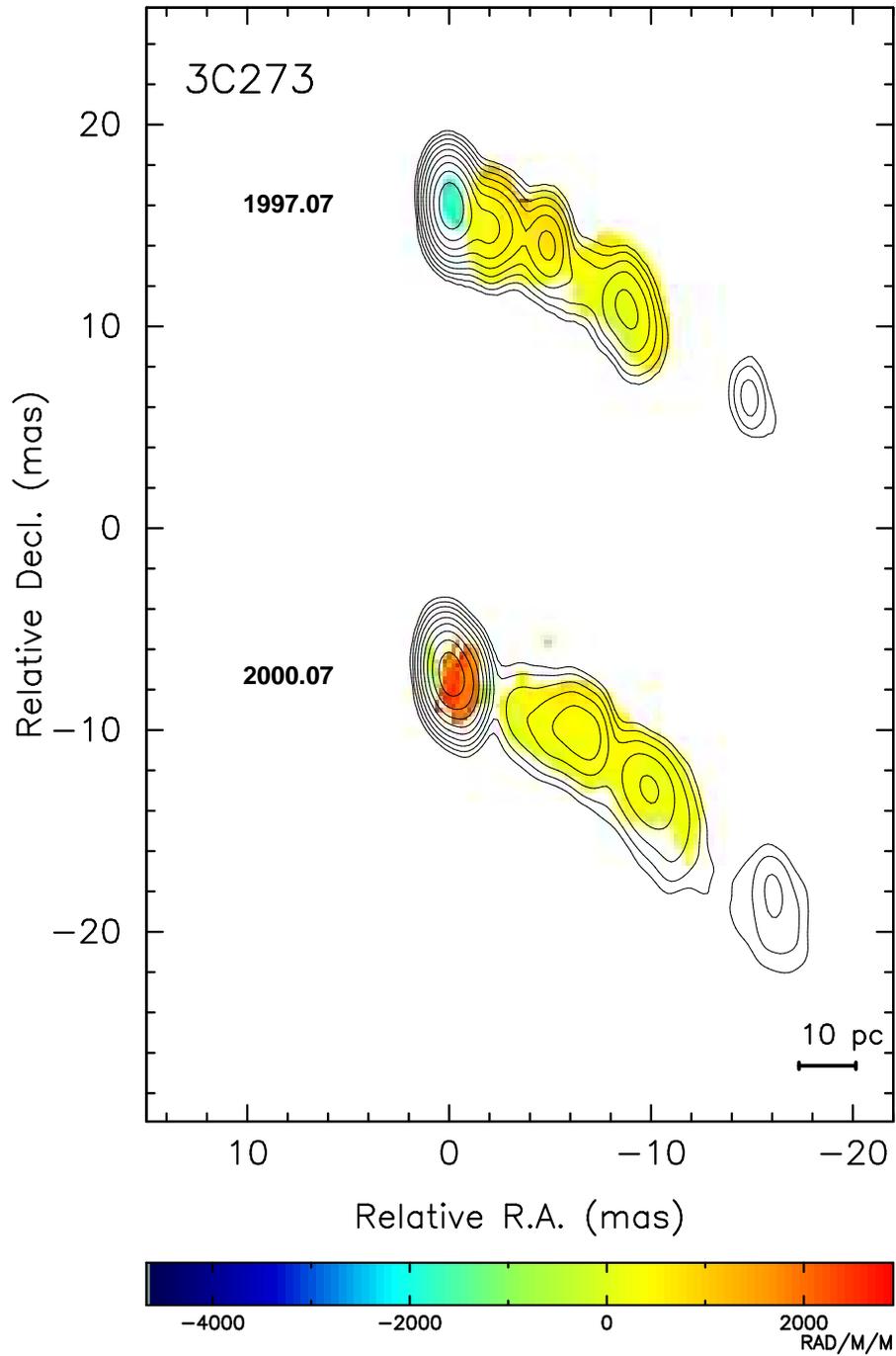}
\figcaption{Rotation measure mosaic of 3C\,273 from data at 8 to 15 GHz for 
the indicated epochs, with contours of total intensity at 8 GHz overlaid.  The 
restoring beam has dimensions 1.0 $\times$ 2.5 milliarcsec at position angle 
0\dg.
The colorbar ranges from $-$4500 to $+$3000 \radm.
Contours start at 30 mJy/beam and increase by factors of 2.}
\end{figure}
\clearpage

\begin{figure}
\vspace{20.0cm}
\includegraphics{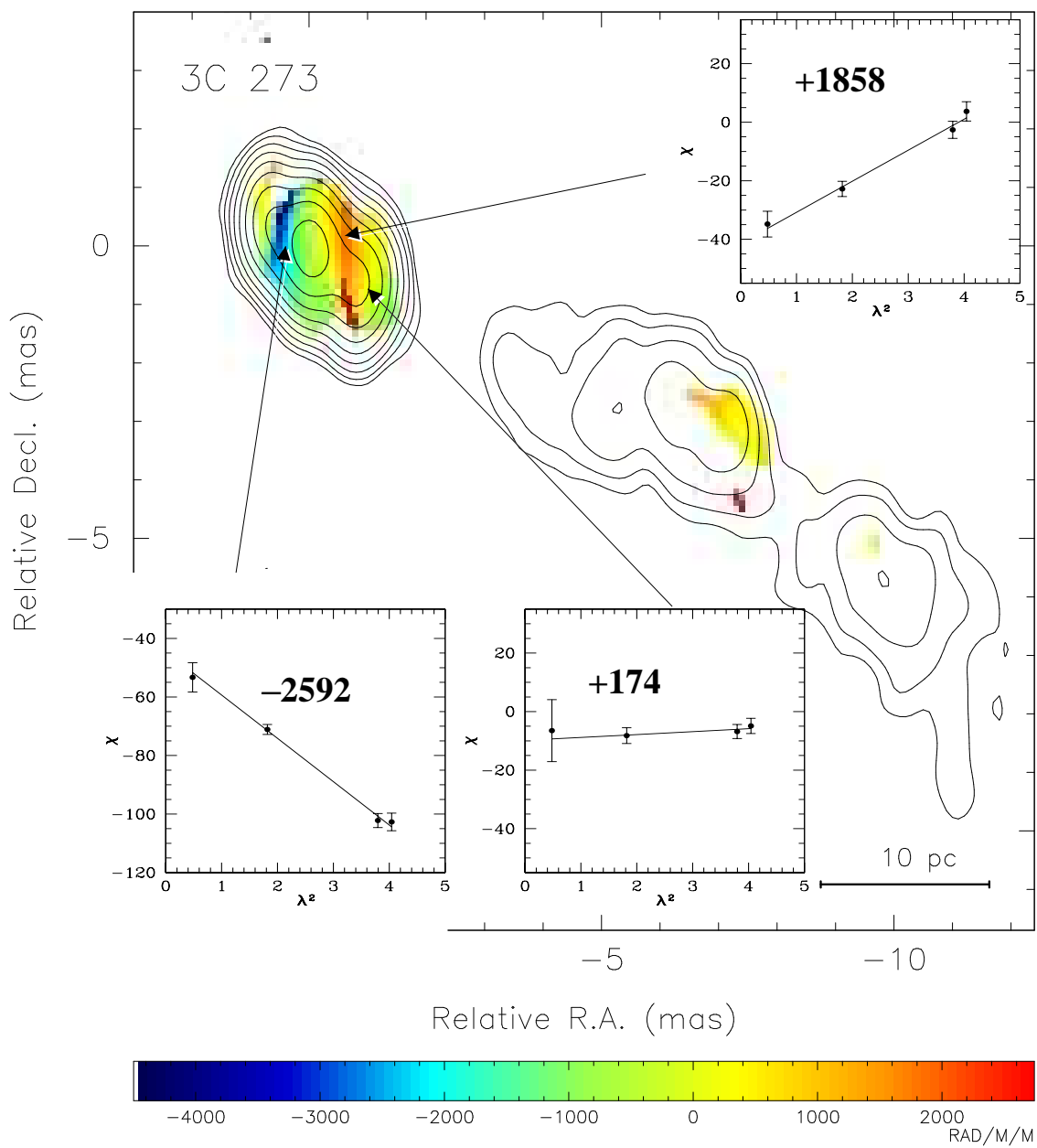}
\figcaption{Rotation measure image of 3C\,273 from 2000.07 data at 43 to 15 GHz,
with contours of total intensity at 22 GHz overlaid. The insets display 
rotation measure fits at the points indicated in \radm. The restoring beam has
dimensions of 0.6 $\times$ 1.2 milliarcsec at position angle 0\dg. The 
colorbar ranges from $-$4400 to $+$2650 \radm. Contours start at 20 mJy/beam 
and increase by factors of 2.}
\end{figure}
\clearpage

\begin{figure}
\vspace{19.2cm}
\includegraphics{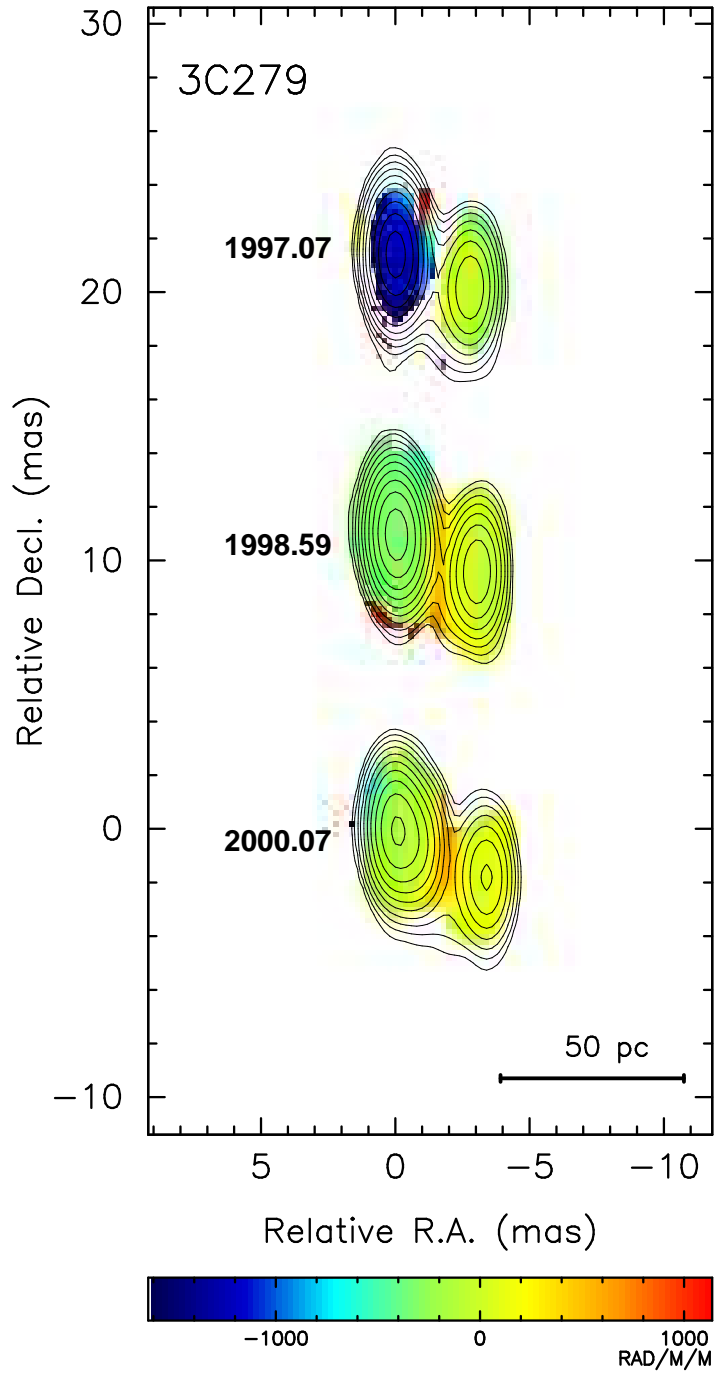}
\figcaption{Rotation measure mosaic of 3C\,279 from data at 8 to 15 GHz, for the
indicated epochs, with contours of total intensity at 15 GHz overlaid. The 
restoring beam has dimensions 1.0 $\times$ 2.5 milliarcsec at position angle 0\dg. The 
colorbar ranges from $-$1310 to $+$1000 \radm.
Contours start at 25 mJy/beam and increase by factors of 2.}
\end{figure}
\clearpage

\begin{figure}
\vspace{20.0cm}
\includegraphics{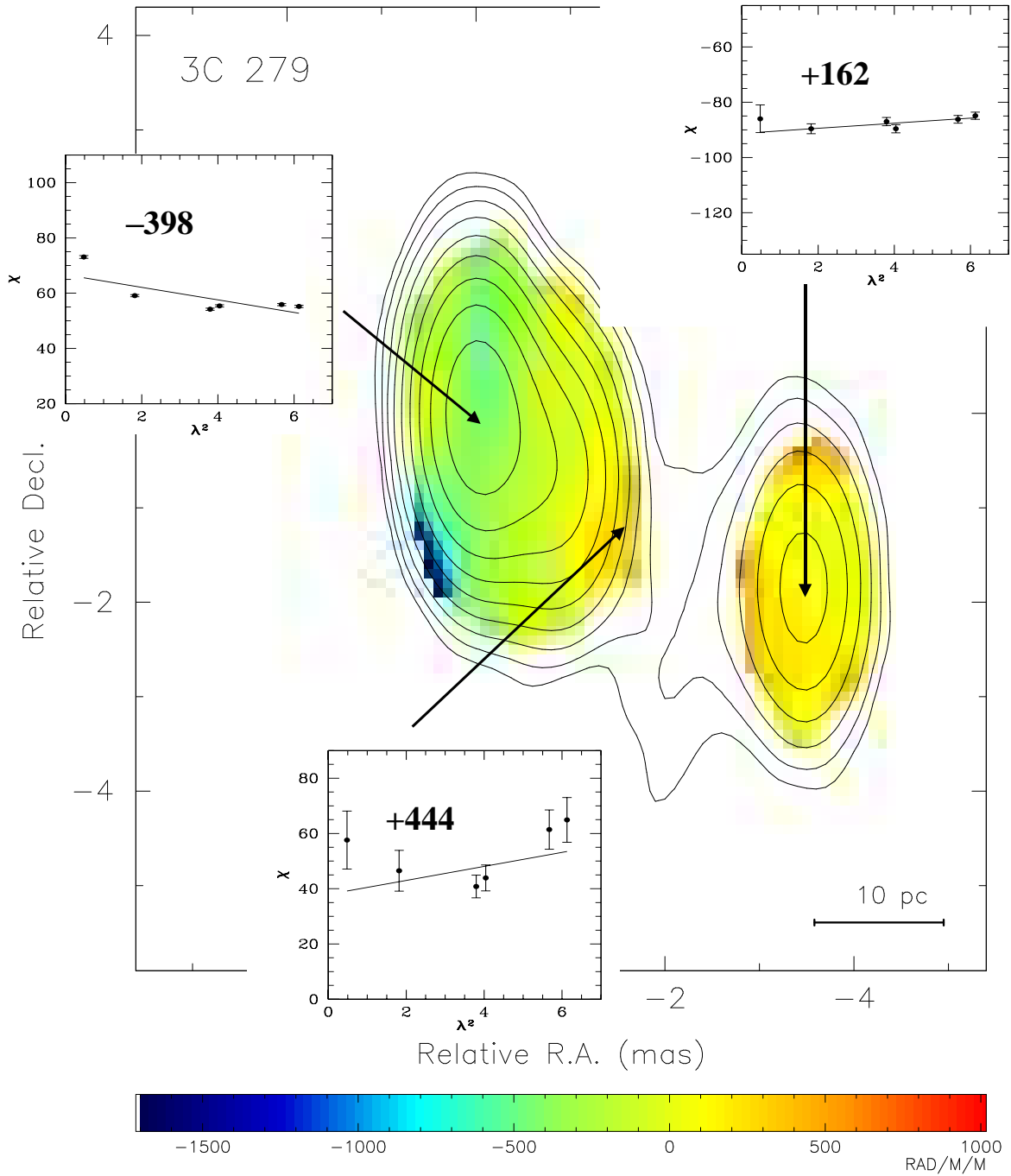}
\figcaption{Rotation measure image of 3C\,279 from data at 12 to 43 GHz, 
with contours of total intensity at 22 GHz tapered to 12 GHz resolution 
overlaid. The insets display rotation measure fits at the points indicated in 
\radm. The restoring beam has dimensions 0.7 $\times$ 1.7 milliarcsec with 
bpa 
0\dg. The colorbar ranges from $-$1000 to $+$1000 \radm.
Contours start at 30 mJy/beam and increase by factors of 2.}
\end{figure}
\clearpage

\end{document}